\input mn.tex
\input epsf.tex

\hyphenation{Pij-pers}
\hyphenation{mat-ri-ces}
%
\newcount\eqnumber
\eqnumber=1
\newcount\apeqnumb
\apeqnumb=1
\def\neqn{{\rm(\the\eqnumber)}\global\advance\eqnumber by 1}
\def\neqap{{\rm(A\the\apeqnumb)}\global\advance\apeqnumb by 1}
\def\refeq#1){\advance\eqnumber by -#1 {\rm(\the\eqnumber)} \advance
\eqnumber by #1}
\def\eqnam#1#2{\immediate\write1{\xdef\ #2{(\the\eqnumber}}\xdef#1{(\the\eqnumber}}
\def\apeqnam#1#2{\immediate\write1{\xdef\ #2{(A\the\apeqnumb}}\xdef#1{(A\the\apeqnumb}}
\newcount\fignumber
\fignumber=1
\def\nfig{\global\advance\fignumber by 1}
\def\refig#1{\advance\fignumber by -#1 \the\fignumber \advance\fignumber by #1}
\def\fignam#1#2{\immediate\write1{\xdef\ #2{\the\fignumber}}\xdef#1{\the\fignumber}}
\def\m@th{\mathsurround=0pt}
\def\twomat#1{\vcenter{\normalbaselines\m@th
    \ialign{$##\hfil$&\quad##\hfil\crcr#1\crcr}}}
\def\ref{\par\noindent
	\hangindent=0.7 truecm
	\hangafter=1}
%
%
\def\etal{{et al.}}

\loadboldmathnames

\begintopmatter

\title{Unbiased image reconstruction as an inverse problem }

\author{ F.P. Pijpers }

\affiliation{
Theoretical Astrophysics Center, Institute for Physics and Astronomy, 
Aarhus University, Ny Munkegade, \goodbreak
8000~\AA{}rhus~C, Denmark }

\shortauthor{F.P. Pijpers }
\shorttitle{Unbiased image reconstruction}

\acceptedline{}

\abstract{
An unbiased method for improving the resolution of astronomical images 
is presented. The strategy at the core of this method is to establish
a linear transformation between the recorded image and an improved
image at some desirable resolution. In order to establish this transformation
only the actual point spread function and a desired point spread function 
need be known. Any image actually recorded is not used in establishing 
the linear transformation between the recorded and improved image.

This method has a number of advantages over other methods currently in
use. It is not iterative which means it is not necessary to impose
any criteria, objective or otherwise, to stop the iterations. The
method does not require an artificial separation of the image into
``smooth'' and ``point-like'' components, and thus is unbiased with 
respect to the character of structures present in the image. The method 
produces a linear transformation between the recorded image and the 
deconvolved image and therefore the propagation of pixel-by-pixel flux 
error estimates into the deconvolved image is trivial. It is explicitly 
constrained to preserve photometry and should be robust against random 
errors.
}


\maketitle

\section{Introduction}

In astronomy the problem of correcting images for imperfect telescope
optics, atmospheric turbulence and other effects that adversely
influence the image quality is very well known. There exist therefore
many different ways to improve on images or imaging techniques in 
order to obtain more detailed spatial information.
First of all there are hardware based solutions such as adaptive optics
and interferometry. It is of course always desirable to make use
of such techniques whenever possible. However, given a recorded
image with a known point spread function (PSF) and an estimate of
the point by point error it is still possible to improve on the
spatial resolution by means of software~: numerical image deconvolution.

A number of algorithms already exist that attempt to do this 
image reconstruction. Best known in the field of interferometry at radio
wavelengths are probably the CLEAN method (H\"ogbom, 1974~;
Schwarz, 1978~; Wakker \& Schwarz, 1988) and the maximum entropy
method (MEM) (cf. Narayan \& Nityananda, 1986). For the deconvolution
of optical images the Richardson-Lucy (RL) method is well known
(cf. Richardson, 1972~; Lucy, 1974, 1992, 1994). Quite recently
a new method was presented by Magain, Courbin, \& Sohy (MCS, 1998).
A characteristic of all of these methods is that images are reconstructed 
by placing flux (MEM and RL), or building blocks such as point sources
(MCS, CLEAN, two channel RL) in the field of view and minimizing
the difference between the image of this model (after convolution with 
the PSF) and the actual image. Since this is an inverse
problem the solution is generally not unique and it can be quite sensitive 
to errors in the data. Thus in the minimization there is some need for 
regularization which is usually a smoothness constraint.
The various method differ in which building blocks are used and in the 
form of regularization applied.

It is however somewhat unsatisfactory to proceed in this manner since
astronomical images are not generally easily described by just
point sources or objects of a certain shape or size, and may
not conform to the smoothness constraint applied. Forcing an
algorithm to nevertheless build the image up within such constraints
may well introduce an undesirable bias. For this reason it is useful to 
consider an alternative that does not assume anything about properties 
of the image in the deconvolution. In fact the method presented here 
does not even use the image actually recorded until its very last step. 

The method of subtractive optimally localized averages (SOLA) was
originally developed for application in the field of helioseismology
(cf. Pijpers \& Thompson, 1992, 1994) to determine
internal solar structure and rotation. Since then it has also been 
successfully applied to the reverberation mapping of the broad line 
region of active galactic nuclei (AGN) (Pijpers \& Wanders, 1994).
Instead of operating on the image, the SOLA method uses the PSF with 
which the image was recorded. With this PSF and a user-supplied desired 
PSF a linear transformation is constructed between any recorded
image for which that PSF applies and its deconvolved counterpart. The 
resolution that can be attained in
this way is only limited by the sampling of the recording device
(the pixel size of the CCD) and by the level of the flux errors in the
recorded image. Since the transformation is linear, it is quite
straightforward to impose photometric accuracy. Astrometric
accuracy at the pixel scale is similarly guaranteed since there is
no `positioning of sources' in the image by the algorithm. Sub-pixel
accuracy, claimed for some deconvolution methods, implies subdividing each
pixel into subpixels. It requires knowledge of the PSF at very high 
accuracy and very small errors in the data in order to deconvolve
down to a sub-pixel scale. If such information is available the
SOLA method can easily accommodate sub-pixel scale deconvolution,
without substantial modifications. In what follows however, it is assumed 
that a single pixel is the smallest scale required.

In section 2 the SOLA method is presented. In section 3 the method
is applied to an example image to demonstrate the workings of
SOLA. In section 4 the method is applied to some astronomical images.
Some conclusions are presented in section 5.

\section{The SOLA method}

\subsection{arbitrary PSFs}

The strategy of the SOLA method in general is to find a set of linear
coefficients $c$ which, when combined with the data, produce a weighted
average of the unknown convolved function under the integral sign, where
the weighting function is sharply peaked. In the application at hand this 
means finding the linear transformation between an image recorded at
a given resolution and an image appropriate to a different (better)
resolution.

The relation between a recorded image $D$ and the actual distribution of
flux over the field of view $I$ is~:
\eqnam\blurone{blurone}
$$
D(x,y)\ =\ \int\ {\rm d}x'{\rm d}y'\ K(x',y'~; x, y) I(x',y') 
\eqno\neqn
$$
where $K$ is the PSF. If one assumes that the PSF is constant over the 
field of view then~:
\eqnam\psfcon{psfcon}
$$
K(x',y'~; x, y) \equiv K (x-x', y-y')
\eqno\neqn
$$
Of course generally $D$ is not known as a continuous function of
$(x,y)$, but instead it is sampled discretely as for instance 
an image recorded on a CCD. Thus one has as available data the recorded
pixel-by-pixel values of flux $D(x_i, y_j)$. These measured fluxes 
will usually be corrupted by noise and thus the discretized version
of equation \blurone) is~:
\eqnam\blurdis{blurdis}
$$
D_{ij}\ =\ \int\ {\rm d}x'{\rm d}y'\ K_{ij}(x',y') I(x',y') + n_{ij}
\eqno\neqn
$$
where now $K_{ij}$ refers to the PSF appropriate for
the pixel at $(x_i,y_j)$ and $D_{ij}$ is the flux value recorded in
that pixel. In the vocabulary usual for the SOLA method the $K_{ij}$
are referred to as integration kernels.

In the SOLA technique a set of linear coefficients ${c_l}$ is sought
which, when combined with the data, produces a value for the flux
$R$ in any given pixel that would correspond to an image recorded
with a much narrower PSF. Writing this out explicitly and using \blurdis) 
yields~:
$$
\eqalign{
R &\equiv \sum c_l D_l = \cr
&\int\ {\rm d}x'{\rm d}y'\ \left\{\sum c_l K_{l}(x',y') 
\right\} I(x',y') + \sum c_l n_{l} \cr}
\eqno\neqn
$$
in which the double subscript $ij$ has been replaced by a single one 
$l$ for convenience. Thus one would construct the ${c_l}$ such that
the averaging kernel ${\cal K}$ defined by~:
$$
{\cal K}(x',y') \equiv \sum c_l K_{l}(x',y')
\eqno\neqn
$$
is as sharply peaked as possible. If one does this for all locations on
the CCD the collected values $R_m$ are then the fluxes
corresponding to the image at this (better) resolution with
a (improved) ``point spread function'' ${\cal K}$. The so-called propagated
error, the error in the flux $R$ is~:
\eqnam\fullerr{fullerr}
$$
\sigma_R^2 \equiv \sum\sum c_l c_m N_{lm} 
\eqno\neqn$$
Here the $N_{lm}$ is the error variance-covariance matrix of the
recorded CCD images where both $l$ and $m$ run over all $(i,j)$ combinations
of the pixel coordinates. If the errors are uncorrelated between pixels 
then \fullerr) reduces to~:
\eqnam\simperr{simperr}
$$
\sigma_R^2 = \sum c_l^2 \sigma_{l}^2
\eqno\neqn$$
which is trivially computed once the coefficients $c_l$ are known.

Ideally one would 
wish to construct an image corresponding to an infinitely narrow
PSF~: a Dirac delta function. In practice this
cannot be achieved with a finite amount of recorded data. As
has already been pointed out by Magain \etal (1998) one must 
in the deconvolved image still satisfy the sampling theorem.
A further restriction arises because of the noise term in equation
\blurdis). As is well known in helioseismology the linear combination
of data corresponding to a very highly resolved measurement usually
bears with it a very large propagated error. In order to obtain
a flux value for each pixel in the deconvolved image that does
not have an excessively large error estimate associated with it, one
needs to remain modest in the resolution sought for in the deconvolved 
image. 

Finding the optimal set of coefficients taking these limitations
into account can be expressed mathematically in the following minimization
problem. One needs to minimize for the coefficients $c_l$ the
following~:
\eqnam\minimize{minimize}
$$
\int\ {\rm d}x{\rm d}y\ \left[ {\cal K} - {\cal T}\right]^2 + \mu \sum
\sum c_{l} c_{m} N_{lm}
\eqno\neqn
$$
Here $\mu$ is a free parameter which is used
to adjust the relative weight given to minimizing the errors in the
deconvolved image and to producing a more sharply peaked kernel
${\cal K}$. The higher the value of $\mu$ the lower this error but
the less successful one will be in producing a narrow PSF.
In SOLA one is free to choose the function ${\cal T}$. A common choice
in SOLA applications is a Gaussian~:
$$
{\cal T} = {1\over f \Delta^2} \exp \left[ - \left({(x-x_0)^2+(y-y_0)^2 
\over\Delta^2}\right)\right]
\eqno\neqn
$$
Here $(x_0, y_0)$ is the location for which one wishes to know the flux
at the resolution corresponding to the width $\Delta$. $f$ is a 
normalization factor chosen such that~:
$$
\int\ {\rm d}x{\rm d}y\ {\cal T} \equiv 1
\eqno\neqn
$$
although any set of locations $(x_0, y_0)$ can be chosen, a natural 
choice in the application at hand is to take all original pixel locations 
$(x_i, y_j)$. If one wishes to deconvolve to sub-pixel scales this
can be done by an appropriate choice of the $(x_0, y_0)$ and $\Delta$.

In terms of an algorithm the problem of minimizing the function \minimize)
leads to a set of linear equations~:
\eqnam\lineqs{lineqs}
$$
A_{lm} c_{l} = b_{m}
\eqno\neqn
$$
The elements of the matrix $A$ are given by~:
\eqnam\matrix{matrix}
$$
A_{lm} \equiv \int {\rm d}x{\rm d}y\ K_l (x,y) K_m (x,y) + \mu N_{lm}
\eqno\neqn
$$
The elements of the vector $b$ are given by~:
\eqnam\vector{vector}
$$
b_{m} \equiv \int {\rm d}x{\rm d}y\ {\cal T} (x,y) K_m (x,y)
\eqno\neqn
$$
Writing out the dependencies on the free parameters explicitly, 
determining the coefficients $c_l$ results from a straightforward
matrix inversion~:
\eqnam\matinv{matinv}
$$
c_l (x_0, y_0~; \Delta, \mu) = A^{-1}_{lm} (\mu) b_m (x_0, y_0~; \Delta)
\eqno\neqn
$$
It is clear that for each point $(x_0, y_0)$ there is a separate set
of coefficients $c_l$ which will depend on the resolution width $\Delta$ 
required and on the error weighting $\mu$.
Note that it is not necessary to invert a matrix for every location
$(x_0, y_0)$, which would certainly be prohibitive if one wishes to
calculate the entire deconvolved image. For a given error weighting
$\mu$ one needs to invert $A$ only once. Only the elements of the
vector $b$ need be recomputed for different locations or different
resolutions.

In order to ensure that at every point in the reconstructed image the
summed weight of all measurements is equal and thus a true (weighted) 
average it is necessary to additionally impose the condition~:
\eqnam\unitav{unitav}
$$
\sum c_l \equiv 1
\eqno\neqn
$$
It is this condition that imposes photometric accuracy on the reconstructed
image. Using the method of Lagrange multipliers this condition is easily
incorporated into the matrix equation \lineqs) by augmenting the
matrix $A$ with a row and column of $1$'s, and a corner element
equal to $0$. The vector $b$ gains one extra element equal to $1$
as well. The details of this procedure can be found in Pijpers \& Thompson
(1992, 1994).

\subsection{translationally invariant PSFs}

Although the method described above can work in principle with general
PSFs $K$, the matrix inversion becomes intractable
very quickly as the number of pixels increases. For an image of
$M \times M$ pixels the number of elements in the matrix A is 
$M^2 \times M^2$. The matrix $A$ is symmetric but even so a naive
matrix inversion routine would require a number of operations
scaling as $M^6$. 

However the entire procedure for obtaining the transformation coefficients 
for all locations of the CCD can be speeded up considerably if one 
accepts some restrictions for the properties of the PSF $K$ and of the
expected errors $N_{lm}$. The first restriction is to assume that the PSF is 
constant over the field of view, that is to say that equation \psfcon) 
is valid. 
When condition \psfcon) is met one can easily demonstrate that 
in equation \matrix) the integrals of the cross products of the PSFs 
are a convolution~:
\eqnam\matconvlv{matconvlv}
$$
\eqalign{
\int &{\rm d}x{\rm d}y\ K_l (x,y) K_m (x,y) \equiv \cr
&\int {\rm d}x{\rm d}y\ K(x-x_{i_l}, y-y_{j_l}) K(x-x_{i_m}, y-y_{j_m}) \cr
= &\int {\rm d}x'{\rm d}y'\ K(x', y') K'(\Delta x_{i_m\, i_l}-x', 
\Delta y_{j_m\, j_l}-y') \cr}
\eqno\neqn$$
in which~:
$$
\eqalign{
K'(x-x', y-y') &\equiv K(x'-x, y'-y) \cr
\Delta x_{i_m\, i_l} &\equiv x_{i_m}-x_{i_l}\cr
\Delta y_{j_m\, j_l} &\equiv y_{j_m}-y_{j_l}\cr}
\eqno\neqn$$
Evaluating all the $M^4$ elements of the matrix $A$ is much simplified
by doing this two-dimensional convolution as a multiplication in the 
Fourier domain. This calculation is then dominated by the FFT calculation
which requires ${\cal O}( M^2 \log (M) )$ operations. Similarly the
vectors $b$ in \vector) can be evaluated for all locations $(x_i, y_j)$
with a single two-dimensional convolution of $K$ and $T$, again dominated
by the FFT.

If the CCD pixels are assumed to be equally spaced the matrix $A$ for
$\mu = 0$ can be constructed in such a way that it becomes of a special 
type known as symmetric block circulant with circulant blocks (BCCB),
for which very fast inversion algorithms exist. 
Circulant matrices have the property that every row is identical to the 
previous row, but shifted to the right by one element. The shifting is
`wrapped around' so that the first element on each row is equal to the 
last element of the previous row. Thus the main diagonal elements are 
all equal and on every diagonal parallel to the main diagonal of the 
matrix all elements are equal as well. A BCCB matrix is a matrix that 
can be partitioned into blocks in such a way that each row of blocks is 
repeated by shifting (and wrapping around) by one block in the subsequent 
row of blocks and each individual block is circulant. 
It can be shown that circulant matrices can be multiplied and inverted 
using Fourier transforms, and by extension BCCB matrices 
can be multiplied and inverted using two-dimensional Fourier
transforms. The detailed steps of the algorithm are worked out in the appendix. 

\fignam{\testimgs}{testimgs}
\beginfigure*{1}
\epsfysize=16.8cm
\epsfbox{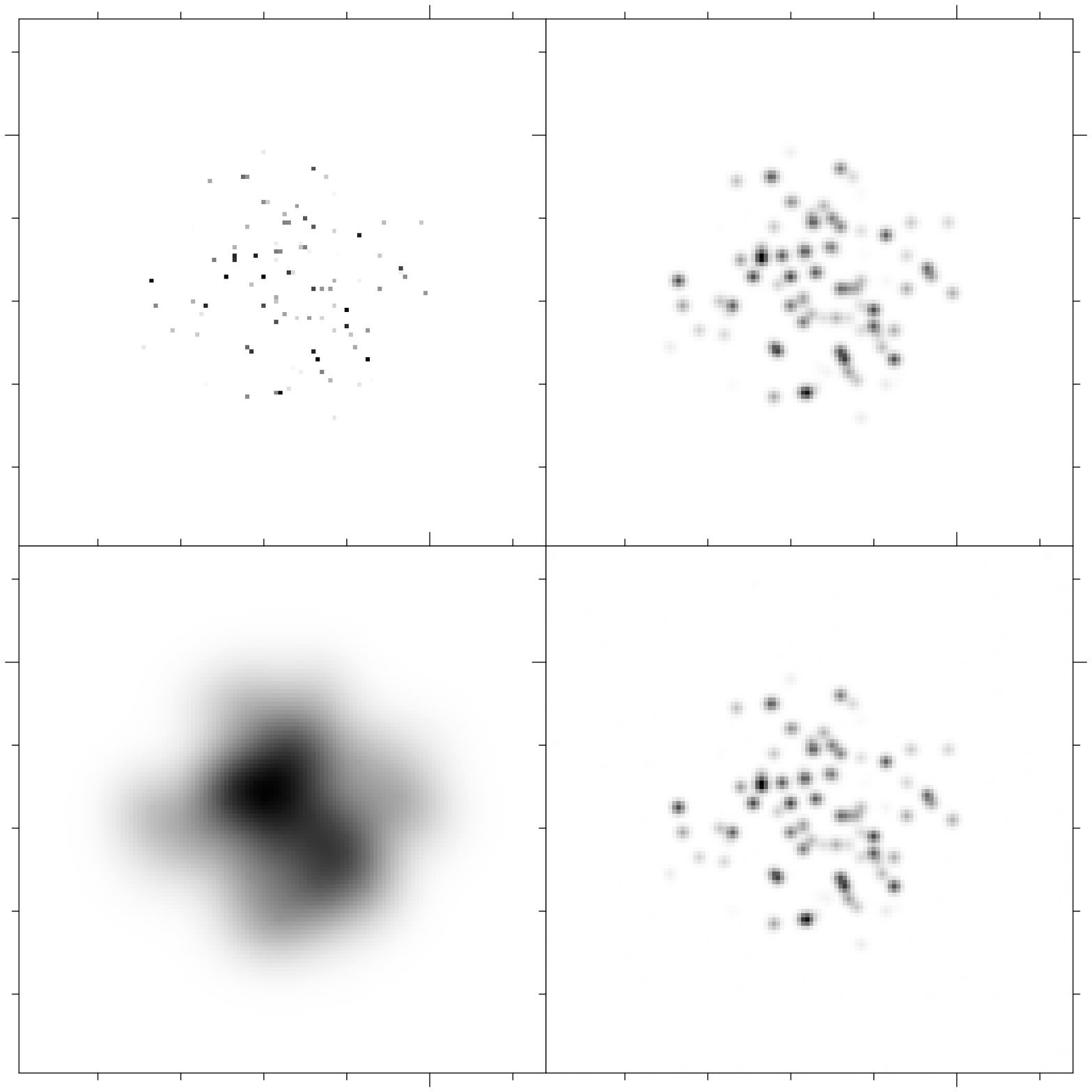}
\caption{{\bf Figure \testimgs.}
{The $128\times 128$ pixels image used in testing the algorithm.
Top left panel~: the original
image. Top right panel~: the original convolved with the target PSF~: 
a Gaussian with $\Delta = 1.5$ pixels. Bottom left panel~: the 
original convolved with a PSF which is the sum of a Gaussian with 
$\Delta = 10$ pixels and an $0.1\%$ contribution from a Gaussian with 
$\Delta = 1$ pixel. 
Bottom right panel~: the image after SOLA deconvolution of the bottom 
left image. In all images the grey-scale is linear. In the
bottom left image noise is added before deconvolution. In the
bottom right image the noise propagated in the deconvolution has
an expectation value of $\sim 0.5$ in arbitrary flux units and the S/N 
ratio for the brightest pixel is $\sim 1000$.
}}
\endfigure
\nfig

The restriction on the matrix $N_{lm}$ is that is must also be a 
symmetric BCCB matrix for the fast inversion algorithm to 
work. It is evident that fully optimal results can only be obtained if 
the full $N^2 \times N^2$ covariance matrix of the errors is used. 
However, the error correlation function for the pixels is expected to 
behave similarly to the point spread function in the sense that it is large 
(in absolute value) for small pixel separations and small for large pixel
separations, independently of where on the CCD the pixel is located. It 
is therefore likely that the error covariance matrix will already be 
BCCB or be very nearly so. Since its role in the minimization of \minimize) 
is to regularize the inversion it is in practice not essential that the 
exact variance-covariance matrix be used. Experience in using SOLA in 
other fields has shown that the results of linear inversions are robust 
to inaccuracies in the error matrix, as long as those are not orders of 
magnitude large~:
if for instance substantial amounts of data (fluxes in pixels) are to 
be given small weight in the resulting linear combination, because of 
large errors associated with them, this can give rise to large departures 
from BCCB behaviour of the error covariance matrix. This would then cause 
problems for the fast version of the SOLA method presented here.
Thus if $N_{lm}$ is not circulant it should in
most cases be sufficient to use a BCCB matrix that is close to the 
original~: one could think of using a modified matrix $\overline{N}_{lm}$ on 
the diagonals of which are the average values over those diagonals of 
the true $N_{lm}$. 
Of course once the coefficients have been determined, when calculating the 
propagated errors one should use equation \fullerr) with the proper 
variance-covariance matrix $N_{lm}$.

As is shown in the appendix the matrix corresponding to the collection 
of all vectors of coefficients $c$, which results from the multiplication 
of $A$ with the matrix corresponding to the collection of all vectors 
$b$ (one for every pixel), is also a BCCB matrix. 
The process of combining these coefficients $c_{l}$ with the recorded 
fluxes on the CCD to form the improved image is~:
$$
R_{ij} = \sum\limits_{k}\sum\limits_{l} C_{kl} D_{i+k-1\, j+l-1}
\eqno\neqn
$$
Since the matrix $C$ is a BCCB matrix and therefore its transpose $C^T$ is
as well, the following holds~:
\eqnam\imaconv{imaconv}
$$
\eqalign{
R_{ij} &= \sum\limits_{k,\, l} C_{kl} D_{i+k-1\, j+l-1}
= \sum\limits_{k,\, l} C^T_{lk} D_{i+k-1\, j+l-1}\cr
&= \sum\limits_{k,\, l} C^T_{2-k\, 2-l} D_{i+k-1\, j+l-1}\cr
&= \sum\limits_{k',\, l} C^T_{k'\, l'} D_{i+1-k'\, j+1-l'}\cr
}\eqno\neqn
$$
From the final equality in \imaconv) it is clear that the process of combining
the matrix of coefficients with the image is a convolution, and 
hence can also be done using FFTs. 

From the above it is clear that limiting the algorithm to the case of 
a PSF that is constant over the CCD implies a profound reduction of the 
computing time. If the PSFs $K$ satisfy the condition \psfcon) the 
vectors $b$ collected together for all $(x_0, y_0) = (x_i, y_j)$
form a BCCB matrix, and therefore the matrix inversion of $A$ {\bf and} 
its subsequent multiplication with {\bf all} vectors $b$, shown in 
equation \matinv), can be done in ${\cal O}( M^2 \log (M) )$ operations. 
The entire deconvolved image is thus produced in ${\cal O}( M^2 \log (M) )$ 
operations. This acceleration of the algorithm over the version described 
in the previous section is so substantial that even when
the PSF is not constant over the CCD it is worthwhile subdividing the
image into subsections in which the PSF can be closely approximated by
a single function $K$. The error introduced in this way can be estimated
in a way similar to what is done in the application of SOLA to the
reverberation mapping of AGN (Pijpers \& Wanders, 1994), and should 
generally be much smaller than the propagated error from equation \fullerr).
If such a subdivision is undesirable, there is the possibility of
reverting to the more general algorithm of section 2.1. For a single
peaked PSF the matrix $A$ should have a banded structure to which
fast sparse matrix solvers can be applied. In this case one could use
the inverse of the matrix $A$ for an approximated PSF that is 
translationally invariant as a pre-conditioner to speed up the matrix 
inversion for the case of the true PSF. 

\section{application to a test image}

\subsection{constructing a narrow PSF}

\fignam{\diffbw}{diffbw}
\nfig
\beginfigure{2}
\epsfysize=8.5cm
\epsfbox{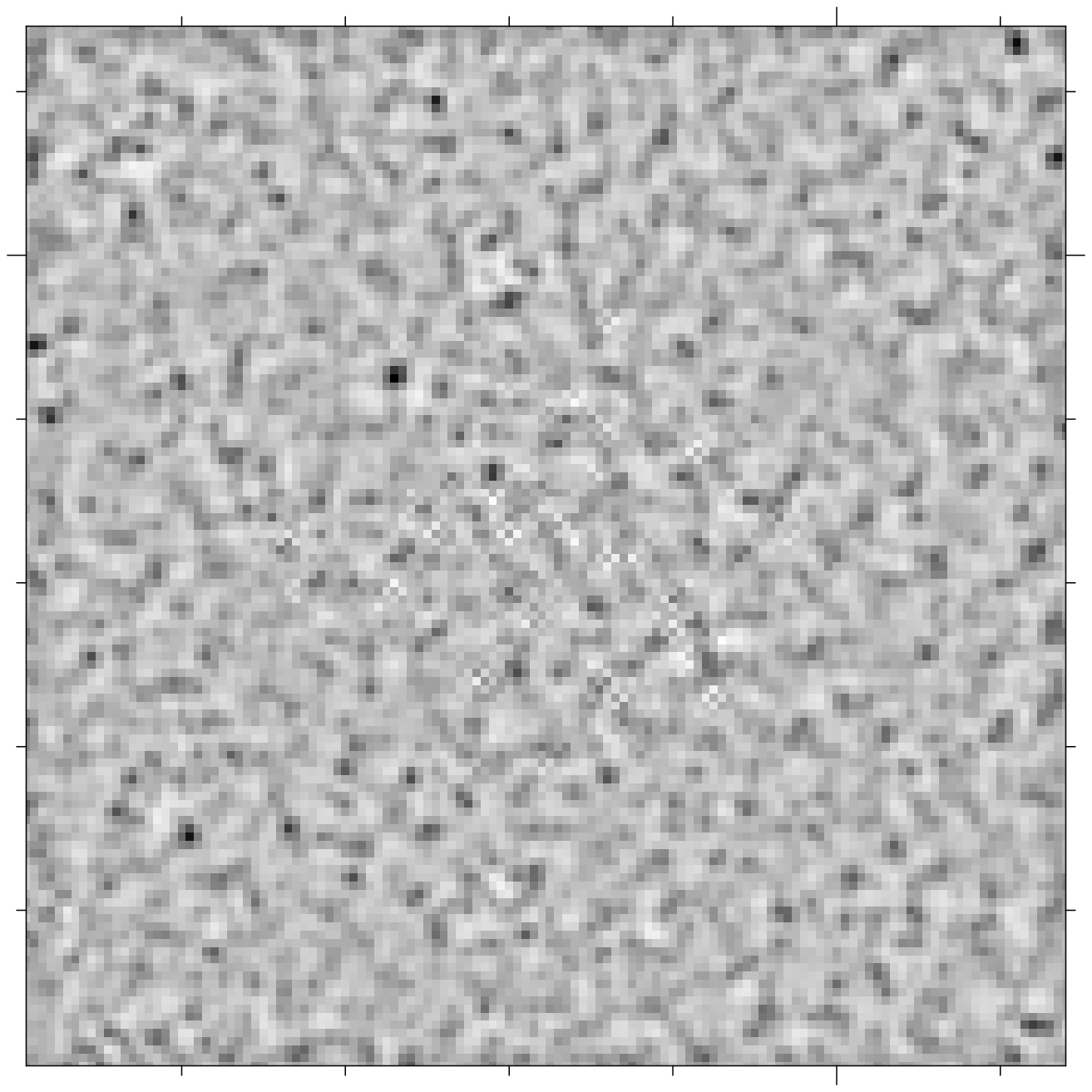}
\caption{{\bf Figure \diffbw.}
{The difference between the image that is SOLA deconvolved
and the original image convolved to the target PSF. 
The gray scale is adjusted so that the full scale is $0.01\times$ the scale 
in the right-hand side images of figure \testimgs{}, which corresponds
to $10\sigma$ of the noise in the deconvolved image.
}}
\endfigure

\fignam{\AvTarker}{AvTarker}
\nfig
\beginfigure{3}
\epsfysize=5.5cm
\epsfbox{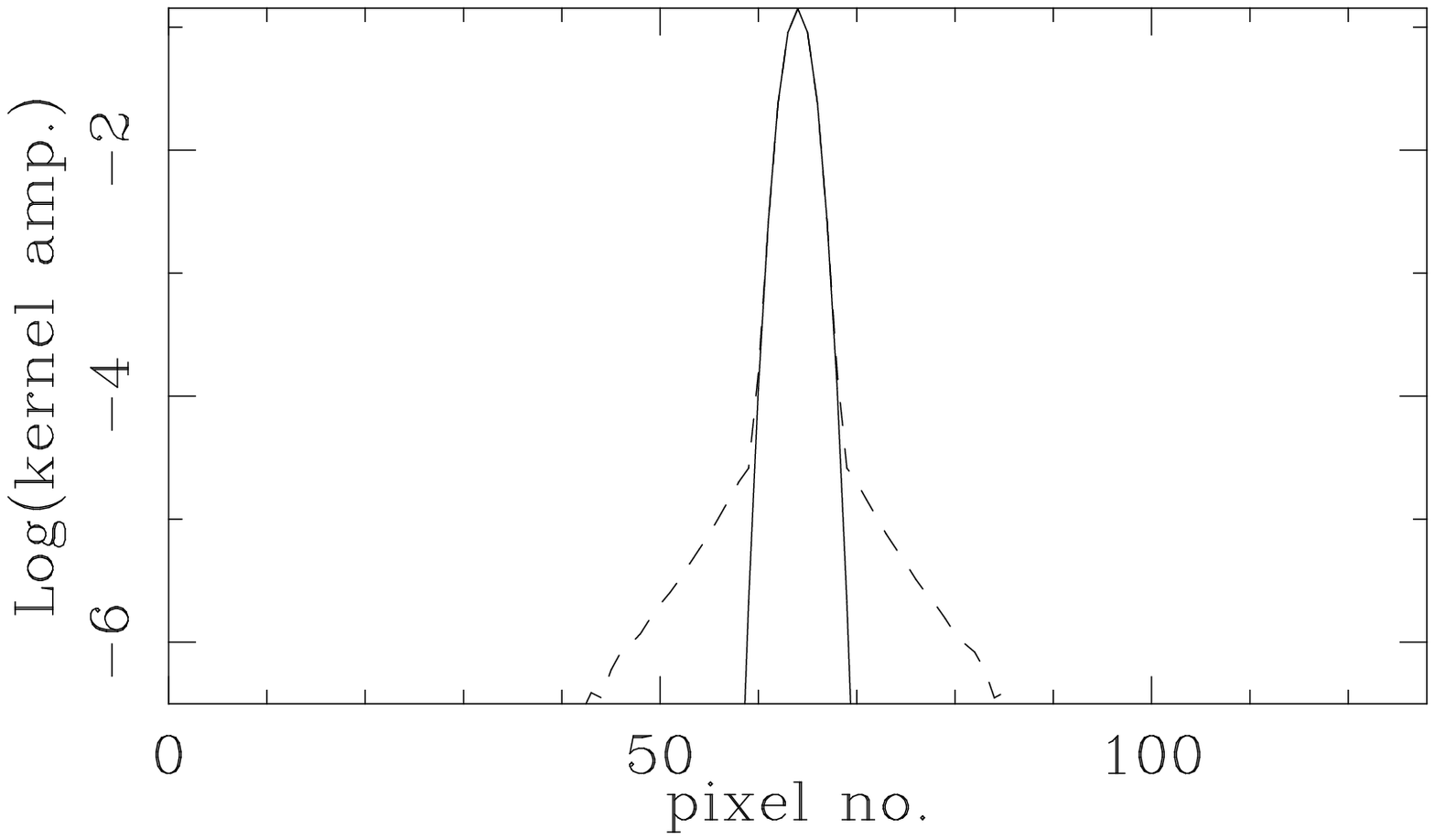}
\caption{{\bf Figure \AvTarker.}
{A slice through the peak of the target PSF specified
in the SOLA algorithm (solid line) and the averaging kernel ${\cal K}$
(dashed) constructed from the linear combination of the pixel PSFs. 
}}
\endfigure

\noindent
In the first instance it is useful to test the algorithm on a
test image for which the result and the errors are known. To this end an 
artificial image of a cluster of stars is convolved with two different 
PSFs. One PSF is a sum of two Gaussians~;
one with a width $\Delta = 10$ pixels in which $99.9\%$ of the flux is 
collected, and a second one with a width $\Delta = 1$ pixels
which collects the other $0.1\%$ of the flux. Poisson distributed noise 
is added to every pixel and this `dirty' image serves as the image to 
be deconvolved.
The other PSF is a Gaussian with a width $\Delta = 1.5$ pixels which 
is also the target chosen for the SOLA algorithm. Thus the deconvolved image
can be compared directly with the image obtained from direct convolution
of the original with the narrow target PSF. The results are shown in
figure \testimgs{}. In order to get an optimal reproduction of the target
PSF, the error weighting parameter $\mu$ is chosen to be equal to $0$. 

\fignam{\DelLam}{DelLam}
\beginfigure{4}
\epsfysize=5.5cm
\epsfbox{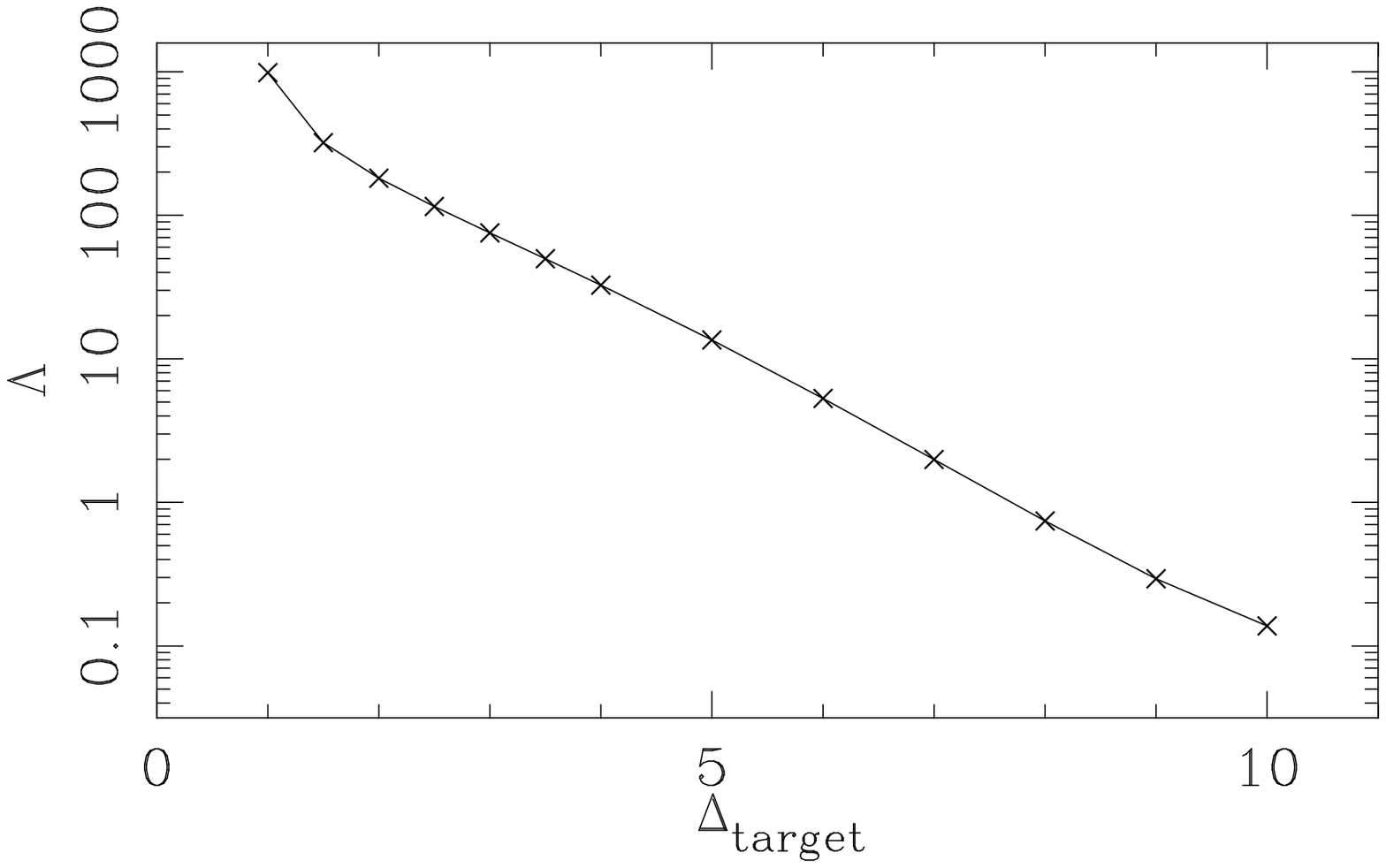}
\caption{{\bf Figure \DelLam.}
{The error magnification $\Lambda$ as a function of the width
$\Delta$ of the target PSF specified. The PSF of the blurred image
is as described in the text in all cases, the error weighting is
$\mu = 0$.
}}
\endfigure
\nfig

It is clear that the bottom and top right panels are very similar 
and thus the image appears to be recovered quite well. 
To illustrate this further the two images can be subtracted.
Figure \diffbw{} shows the SOLA deconvolved image minus the image convolved
with the target PSF, with an adjusted gray scale to bring out the
differences, which in the central portion of the image are all $< 1 \%$. 
Although there is no strong evidence for it in this image, the deconvolution 
can suffer from edge effects because part of the original image can 
`leak away' in the convolution with the broad PSF. When deconvolving,
the region outside the image is assumed to be empty and so a spurious 
negative signature is then introduced in the image. The magnitude of such
edge effects must clearly depend both on the image and on the PSF of
the `dirty' image, since they are determined by the information that has been
lost at the edges of the CCD. Of course it is desirable to demonstrate 
this method on a more realistic suite of images than just the simple 
one used here, which is work currently in progress.

\fignam{\difmag}{difmag}
\beginfigure{5}
\epsfxsize=8.5cm
\epsfbox{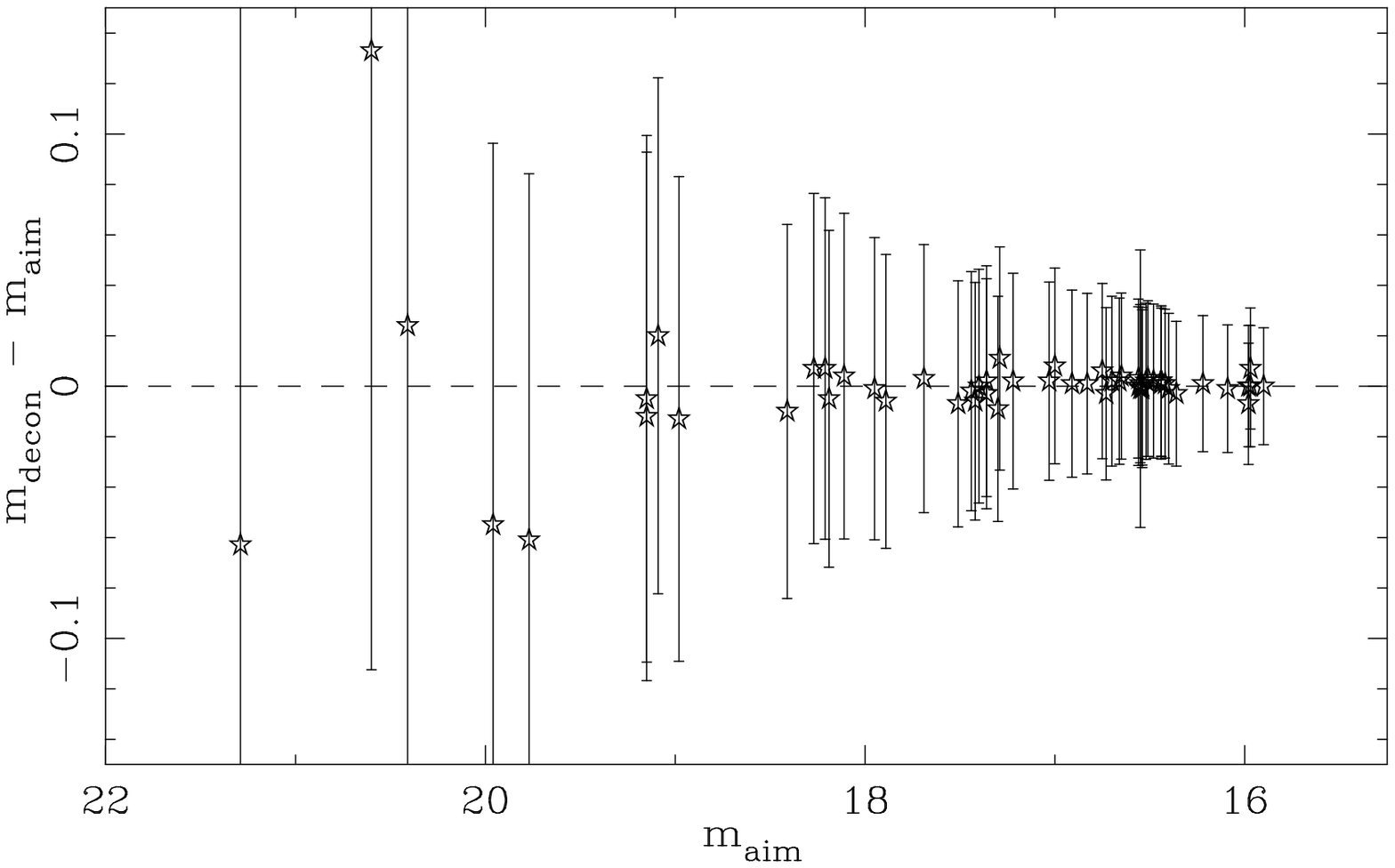}
\caption{{\bf Figure \difmag.}
{The difference on an arbitrary magnitude scale between the magnitude of 
the stars in the deconvolved image $m_{\rm decon}$ and the magnitude
$m_{\rm aim}$ of their counterparts in the reference image constructed by 
convolving the original with the target PSF.
}}
\endfigure
\nfig

\fignam{\difpos}{difpos}
\beginfigure{6}
\epsfxsize=8.5cm
\epsfbox{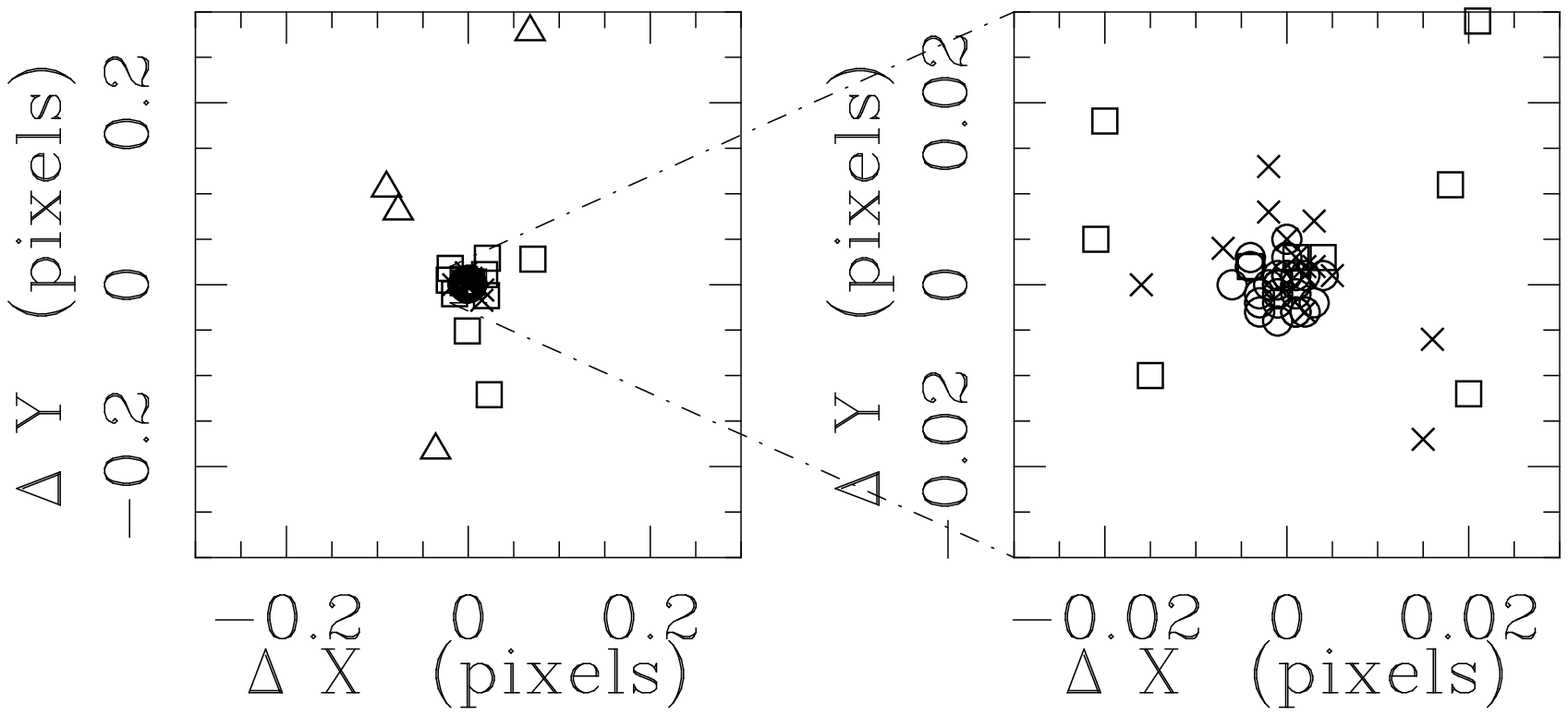}
\caption{{\bf Figure \difpos.}
{The position difference in pixel units between the stars in the deconvolved 
image $m_{\rm decon}$ and their counterparts in the image constructed 
by convolving the original with the target PSF. Open circles : all
stars with magnitudes between $15.9$ and $16.9$, crosses : stars with
magnitudes between $16.9$ and $17.9$, open squares : stars with magnitudes 
between $17.9$ and $18.9$, open triangles : all stars with magnitude 
greater than 18.9.
}}
\endfigure
\nfig

The averaging kernel that is constructed cannot in general match perfectly
the target form, even in the absence of errors. In general any function
can be completely reconstructed only out of a {\it complete set} of base 
functions. Since function space is infinite dimensional this would 
require an infinite number of base functions. In this test image there 
are no more available than the $128^2$ PSFs corresponding to each of the 
pixels and so there can never be a perfect matching of ${\cal K}$
with ${\cal T}$. In figure \AvTarker{} a section through the maximum of 
both ${\cal T}$ and ${\cal K}$ is shown. It is clear that at the $10\ {\rm ppm}$
level, the constructed averaging kernel starts getting wider than the target.
If the ratio of the widths of the target form and actual PSF is
even smaller than for this image, alternating negative and positive side 
lobes can show up in the averaging kernel which cause ringing. The
amplitude of the sidelobes, and the width $\Delta$ below which ringing
starts occurring, will in general depend on the weighting of the errors
$\mu$, as has been demonstrated from the application of SOLA to
helioseismology (Pijpers \& Thompson, 1994).

As it stands the SOLA algorithm does not impose positivity on the image. 
One could attempt to use a positivity constraint to extrapolate the image
beyond the recorded edges in such a way that it eradicates any negative
fluxes in the image, which would in principle also remove associated positive 
artifacts around the edges. However, in the presence of errors this might be 
somewhat hazardous. Furthermore, in the presence of errors any 
edge effects might well disappear into the noise. 

If one assumes that
the covariance of errors between pixels is equal to $0$ and the flux
error in each pixel is equal to $\sigma^2$, or $N \equiv \sigma^2 I$, then
it is particularly simple to calculate the flux error for each pixel
in the deconvolved image from equation \simperr) since it is 
$$
\sigma_R^2\ =\ \sigma^2 \sum c_l^2\ \equiv\ \Lambda^2 \sigma^2
\eqno\neqn
$$ 
This factor $\Lambda$ is usually referred to as the error magnification 
and is equal for all pixels in the reconstructed image. In general $\Lambda$
increases as the ratio of $\Delta_{\rm target}/\Delta_{\rm PSF}$
decreases. For the deconvolved image of figure \testimgs{} the error 
magnification is $\sim 321$. 
The magnitude of the error magnification for this simple example 
illustrates that the true limitation of deconvolving images may 
in practice not lie in the sampling theorem, but instead in the
S/N of the recorded image. For example for a point source the 
peak flux in its central pixel will increase in the deconvolution by a factor
which is roughly ${\rm FWHM}_{\rm PSF}/{\rm FWHM}_{\rm target}$. The noise
will increase by a factor $\Lambda$ and so the signal-to-noise
ratio for point sources will scale roughly as~:
$$
\left({S\over N}\right)_{\rm decon} \approx {1\over\Lambda} 
\left({ {\rm FWHM}_{\rm PSF} \over {\rm FWHM}_{\rm target}}\right) 
\left({S\over N}\right)_{\rm dirty}
\eqno\neqn$$
Thus in the example shown the signal-to-noise ratio for point sources
is degraded by a factor of roughly $\sim 48$ between the dirty image
and the deconvolved image. This clearly requires a very high a 
signal-to-noise ratio in the dirty image, which means that in practice
as dramatic a resolution enhancement as attempted here will not
usually be possible. A more modest resolution enhancement of
around a factor of 2 should in most cases be possible however,
as can be deduced from figure \DelLam{}.

In order to show the relation between resolution and error
magnification in figure \DelLam{} is shown the value of the error 
magnification as a function of the width $\Delta$ specified for the 
target function.
If only the broad component had been present the error magnification 
would have been unity for a target $\Delta = 10$. Effectively
because of the narrow component which captures a mere $0.1\%$ of the flux
the image can be deconvolved to a PSF with $\Delta = 8$ pixels 
without significant penalty in the magnification of the errors. 

The CPU time used to construct the matrix $A$, all the vectors $b$, 
to invert $A$ and multiply with all $b$ and finally to combine the
coefficients with the image takes $\sim 0.5\ {\rm min}$ on an SGI
workstation for this $128\times 128$ image.

\subsection{photometric and astrometric accuracy}

Since there is no placement of flux or point sources, the algorithm should
automatically be astrometrically accurate. Photometric accuracy is ensured 
by explicitly constraining the linear coefficients to sum to unity, i.e.
imposing constraint \unitav). In order to demonstrate both these properties
for this test image a standard photometric package DAOPHOT was used to
do aperture photometry on the deconvolved image, and on the image obtained by
convolving the original with the target PSF. The deconvolved image of
course has noise propagated from the dirty image. The other image is
kept noise-free to properly serve as a reference. The errors are calculated
by DAOPHOT and are consistent with what is expected from the noise in 
the deconvolved image. In figure \difmag{} is shown the difference
between the magnitudes of the stars in the two images as a function of 
the stellar magnitudes in the reference image. The difference is clearly 
consistent with zero over the entire range of 5 magnitudes, and does not
show any trend. 

DAOPHOT calculates the error bar assuming that the error in different
pixels is uncorrelated. For the SOLA deconvolved image this is not
the case. In the deconvolved image the error correlation function
falls below $0.01$ in absolute value only at inter-pixel distances
larger than $\sim 12$ for this test case. Furthermore, because fluxes
are combined with positive and negative coefficients, the error
is not distributed as for a Poisson process. If this is taken
into account properly the error bars in figure \difmag{} should be 
decreased and are then compatible with the actual scatter of the points.

\fignam{\sectst}{sectst}
\beginfigure{7}
\epsfxsize=6.8cm
\epsfbox{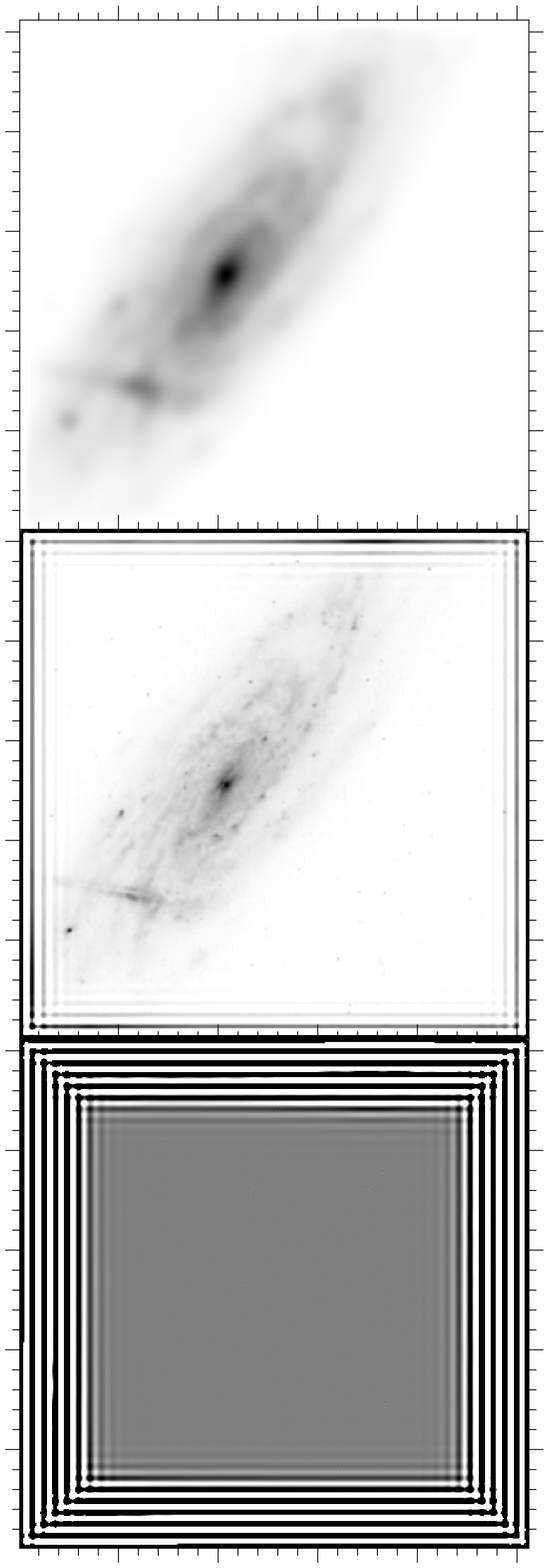}
\caption{{\bf Figure \sectst.}
{Images of the galaxy UGC 5041. The top image is convolved with the same 
broad PSF used on the image at the bottom left in figure \testimgs{}. 
The middle image is the deconvolved image with a PSF with a FWHM of $2.5$. The
bottom image is the difference between the original image convolved
with the target PSF, and the deconvolved image, no noise has been added.
The gray scale of the bottom panel extends between $\pm 0.1\%$
of the gray scale of the middle image.
}}
\endfigure
\nfig

\fignam{\omcdet}{omcdet}
\beginfigure*{8}
\epsfxsize=18.0cm
\epsfbox{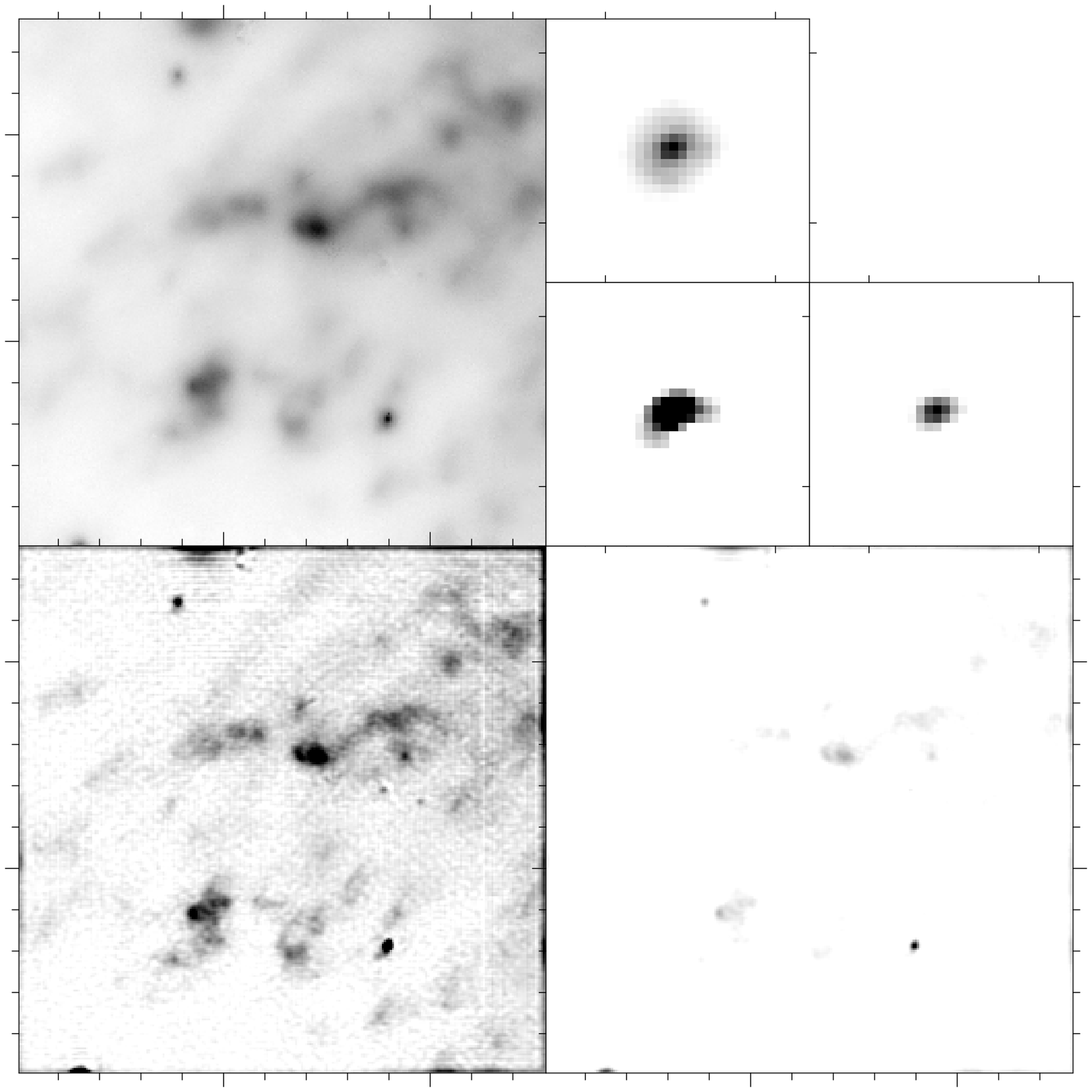}
\caption{{\bf Figure \omcdet.}
{Image of a detail of the Orion Molecular Cloud obtained using adaptive 
optics in IR lines of shocked molecular hydrogen, north is at the top
of the image, east is to the left. The image size is 
$256\times 256$ pixels at $50 {\rm mas/pixel}$. The top left image is 
as obtained with the Adonis instrument on ESO's $3.6{\rm m}$ telescope. 
The PSF for this image is shown on a $4\times$ enlarged scale as the 
$32\times 32$ pixel image in the top left-hand corner of the top 
right-hand panel. The bottom left- and 
right-hand panels are the deconvolved image using different gray-scales. 
The two bottom images in the top right-hand panel show $32\times 32$ 
pixels images of the PSF for the deconvolved image, using the same
gray scales as the corresponding images, and the same spatial scale
as the PSF for the original. For all images 
the dynamic range between lightest and darkest colour is a factor of 
$\sim 8$ in flux level. 
}}
\endfigure
\nfig

DAOPHOT also determines the positions of point sources in the image
and therefore those positions can be used to determine astrometric
accuracy. In figure \difpos{} are shown the difference in units of a
pixel between the DAOPHOT determined positions in the two images.
Here the stars have been grouped into 4 magnitude bins, each
bin 1 magnitude in range, and starting from the brightest star with 
magnitude $15.9$. The right-hand panel is a blow-up of the central
portion of the left-hand panel. Figure 4 shows that there is a trend in
that the fainter stars show a greater scatter in position, the largest 
position difference being of the order of $\sim 0.2$ pixels. In the
right-hand panel it can be seen that for stars brighter than magnitude $19$ 
the difference in positions is smaller than 0.03 pixels. These
uncertainties are entirely consistent with the accuracy with which 
DAOPHOT can determine stellar positions.

From figures \difmag{} and \difpos{} it is clear that if any errors
in photometry or in position are introduced by the deconvolution
process, they are much smaller than the errors due to the random noise.

\section{application to astronomical images}

\subsection{UGC 5041}

To give a somewhat more interesting example, a high resolution HST image
of a galaxy is blurred and then deconvolved to demonstrate that the
method also works on an image which contains a combination of extended
structure and point sources.

The galaxy UGC~5041 is an Sc type galaxy at a redshift of 0.027 (Haynes
\etal., 1997). It has been part of various surveys for use in studies 
of clustering and in establishing distance scales for the Tully-Fisher 
distance method. In figure \sectst{} is shown a $512 \times 512$ image 
of this galaxy obtained in March 1997 using the WFPC2 (WF3) instrument 
on board the Hubble Space Telescope (HST) with the F814W filter which 
corresponds to the I-band. The resolution of the original image has
a FWHM of $1.4$ pixels and is convolved with the same broad PSF used
in the bottom left panel of figure \testimgs{}. It is then deconvolved
to the same resolution as the right-hand images of figure \testimgs{}
and the result is shown in the middle panel of figure \sectst{}.
The difference image between this deconvolved image and a reference 
is shown in the bottom panel of figure \sectst{}, where the gray scale
is enhanced to demonstrate that in the absence of noise the differences
between the deconvolved image and the reference image are less than
$0.1\%$ of the peak flux apart from edge effects.

\subsection{Orion Molecular Cloud}

Observations in IR lines of shocked ${\rm H}_2$ of the SE part of the 
Orion Molecular Cloud complex (OMC1) have been performed at the ESO 3.6m 
telescope taking advantage of the high spatial resolution given by 
adaptive optics (Adonis at 50 marcsec/pixel) combined with the high 
spectral resolution given by a Fabry-Perot (R=1000) (Vannier \etal, 1998).
The image is deconvolved using a target PSF with FWHM $3$ pixels
and an error weighting $\mu = 0$. The resulting error magnification
factor is $\Lambda = 6.2$. The dynamic range in the deconvolved image
is $\sim 30$ as opposed to $\sim 8$ in the original. For this
reason the deconvolved image is shown twice~: bottom left in figure \omcdet{} is 
shown a linear gray scale extending from an estimated noise level to 
$8$ times that, bottom right is shown a linear gray scale extending
from the peak level in the deconvolved image which is $\sim 30$ times
the estimated noise level, to $1/8$ of that. The PSF of the deconvolved
image, i.e. the constructed averaging kernel ${\cal K}$, is shown
using the same two gray scales in the bottom part of the 
top right-hand panel of figure \omcdet{}.

Although some of the `graininess' in the bottom panels must be due to 
the increased noise compared to the top left-hand image, fine structure 
can clearly be seen in the bottom panels of figure \omcdet{}. There
is also some evidence of edge effects at the top and right of the image.
From the image of the PSF of the deconvolved image it is also clear that,
as expected, the Gaussian target is not reproduced perfectly over the
entire dynamic range of the deconvolved image. Comparing the PSF images
with the same dynamic range from the peak down (top left and bottom right
in the top right-hand panel of figure 8) it is clear that
the PSF is indeed much narrower for the deconvolved image.

\section{Conclusions}

In this paper the SOLA inversion method, well known in helioseismology, is
applied to the reconstruction of astronomical images. It is demonstrated
how a linear transformation is constructed between any image recorded
with a known PSF and its deconvolved counterpart with a different
(narrower) PSF. The method itself uses {\bf only} the PSF and no assumptions
are made concerning what is contained within the image(s) to be deconvolved.
It is furthermore shown that in the case of 
translationally invariant PSFs, a fast algorithm, using ${\cal O} (N\log N)$ 
operations where $N$ is the total number of pixels in the image, can be 
constructed, which allows deconvolution of even $1024 \times 1024$ images 
within half an hour on medium-sized workstations. 

\section*{Acknowledgments}
Steve Holland is thanked for a number of helpful discussions. 
D. Rouan, J.-L. Lemaire, D. Field, and L. Vannier are thanked for making
available their data prior to publication.
The observations of UGC~5041 were made with the NASA/ESA 
Hubble Space Telescope, obtained from the data archive at the Space 
Telescope Science Institute. STScI is operated by the Association of 
Universities for Research in Astronomy, Inc. under NASA contract 
NAS 5-26555. 
The Theoretical Astrophysics Center is a collaboration 
between Copenhagen University and Aarhus University and is funded by 
Danmarks Grundforskningsfonden. 

\section*{References}

\ref
Haynes M.P, Giovanelli R., Herter T., Vogt N.P., Freudling W., Maia M.A.G., 
Salzer J.J., Wegner G., 1997,
{Astron. J.}
113, 1197
\ref
H\"ogbom J.A., 1974, 
{A\&{}A Supp.}
15, 417
\ref
Lucy L.B., 1974,
{Astron. J.}
79, 745
\ref
Lucy L.B., 1992,
{Astron. J.}
104, 1260
\ref
Lucy L.B., 1994,
{A\&{}A}
289, 983
\ref
Magain P., Courbin F., Sohy S., 1998,
{ApJ}
494, 472
\ref
Narayan R., Nityananda R., 1986,
{Ann. Rev. A\&{}A}
24, 127
\ref
Pijpers F.P., Thompson M.J., 1992,
{A\&{}A}
262, L33 
\ref
Pijpers F.P., Thompson M.J., 1994,
{A\&{}A}
281, 231 
\ref
Pijpers F.P., Wanders I., 1994,
{MNRAS}
271, 183 
\ref
Press W.H., Teukolsky S.A., Vetterling W.T., Flannery B.P., 1992,
Numerical Recipes, the art of scientific computing 2$^{nd}$ Ed.,
CUP, Cambridge, 70
\ref
Richardson W.H., 1972,
{J. Opt. Soc. Am.}
62, 55
\ref
Schwarz U.J., 1978,
{A\&{}A}
65, 345
\ref
Vannier L., Lemaire J.-L., Field D., Rouan D., Pijpers F.P., Pineau des 
For\^e{}ts G., Gerin M., Falgarone E., 1998,
{ESO/OSA Meeting on Astronomy with Adaptive Optics. Present results and
future programs, Sonthofen Germany 7-11 sept 98},
in press 
\ref
Wakker B.P., Schwarz U.J., 1988,
{A\&{}A}
200, 312

\appendix
\section{The inversion and multiplication of circulant matrices}

As it stands the matrix described in \matconvlv) does not conform to
the criteria for a block circulant with circulant blocks (BCCB) matrix. Instead 
it is a block Toeplitz matrix with Toeplitz blocks. For the latter type 
of matrix the elements are also identical on diagonals but there is no
`wrapping around' from row to row~: the final element of each row is not 
(necessarily) equal to the first element of the subsequent row. 
In order produce a matrix that is a BCCB matrix 
it is useful to envisage a `virtual CCD' that is twice as big in both
dimensions as the actual CCD, which for convenience is assumed to be 
square. This `virtual CCD' has the property that it has a periodic point 
spread function in both directions~: the CCD is shaped like a torus. 
The actual CCD then occupies one quarter of this virtual
CCD and the other three quarters are `empty sky'.

For this virtual CCD the matrix constructed by \matconvlv) is a 
$(2M)^2\times (2M)^2$ BCCB matrix. The first quadrant of this matrix,
$(M)^2\times (M)^2$ in size, is identical to the matrix for the actual
CCD. Unless explicitly stated otherwise the matrices for this torus-shaped 
virtual CCD are the ones that the algorithm works on. In what follows
the indices `wrap around' which is to say that the values of the indices
are to be evaluated modulo the matrix dimension.

First consider a matrix that is fully circulant rather than block
circulant with circulant blocks. Such a matrix $A$ satisfies~:
\apeqnam\fulcirc{fulcirc}
$$
\eqalign{
A_{i+n\, j+n} = A_{ij}\hskip 1cm \forall n,\ &(i+n){\rm mod}(M_A), \cr
&(j+n){\rm mod}(M_A)\cr}
\eqno\neqap
$$
in which the matrix $A$ has dimensions $M_A \times M_A$. 
An ordinary matrix multiplication of two circulant matrices $A$ and $B$ 
satisfies~:
\apeqnam\matmul{matmul}
$$
\eqalign{
C_{ij} &= \sum\limits_k A_{ik} B_{kj} \cr
&= \sum\limits_k A_{i+n\, k+n} B_{k+n\, j+n} \cr
&= \sum\limits_l A_{i+n\, l} B_{l\, j+n} \hskip 1cm {\rm with}\ l\equiv k+n\cr
&=C_{i+n\, j+n} \hskip 1cm\forall n \cr}
\eqno\neqap
$$
Thus $C$ is circulant if $A$ and $B$ are circulant. It is also 
straightforward to demonstrate that if $A$ and $C$ are circulant, $B$ must
be circulant as well~:
$$
\eqalign{
0&= C_{i+n\, j+n} - C_{ij} \cr
&= \sum\limits_l A_{i+n\, l} B_{l\, j+n} - A_{il} B_{lj} \cr
&= \sum\limits_k A_{ik} \left( B_{k+n\, j+n} - B_{kj}\right)\ \ \ k\equiv l-n \cr
}
\eqno\neqap
$$
Since $A$ is an arbitrary circulant matrix the final equality can only
lead to $0$ (and thus consistency) if $B$ is indeed circulant.

Since the identity matrix is itself a circulant matrix, a direct consequence 
is that the inverse of any circulant matrix $A^{-1}$ is also circulant, which is 
trivially demonstrated by substituting $I$ for $C$.

\fignam{\stackblock}{stackblock}
\beginfigure{9}
\epsfysize=5.2cm
\epsfbox{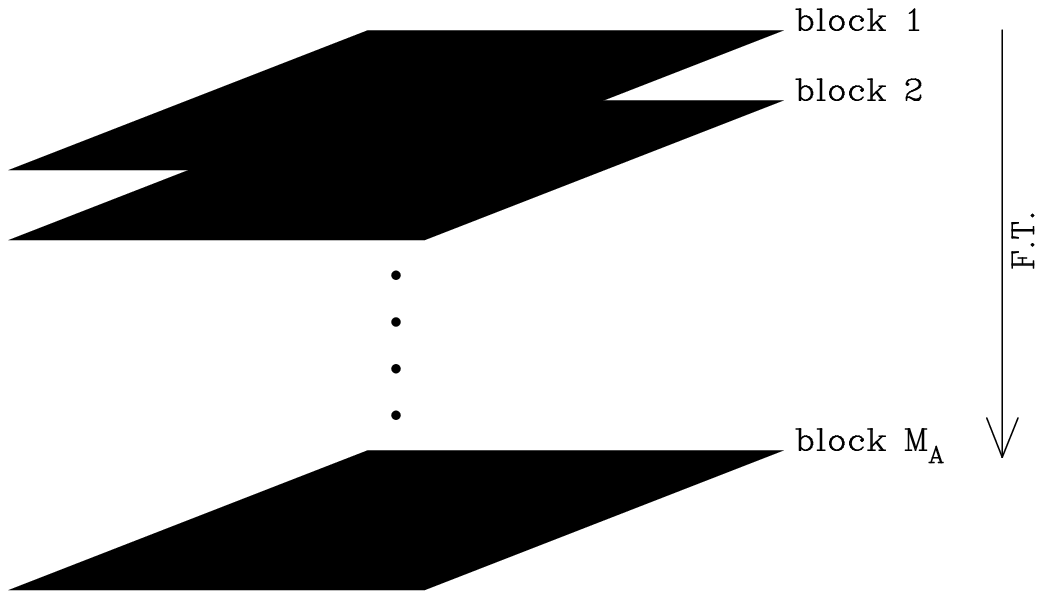}
\caption{{\bf Figure \stackblock.}
{The first row of blocks of a BCCB matrix ar stacked above each
other. The first step in multiplying two BCCB matrices is to
do a FT on these stacks for each matrix in the direction indicated by 
the arrow. Since the blocks individually are circulant one only
needs to do this for the front face of this cube by performing an FT
on each row and multiplying in the Fourier domain. }}
\endfigure
\nfig

If the matrix $C$ is circulant then it is fully determined by its first 
row, as can be demonstrated by taking $n=1-i$ in \matmul), for which the 
following holds~:
\apeqnam\firstrow{firstrow}
$$
\eqalign{
C_{1n} &= \sum\limits_l A_{1l} B_{ln} \cr
&= \sum\limits_l A_{1l} B_{1\, n+1-l} \cr}
\eqno\neqap
$$
A one-dimensional convolution of two functions is defined by
the following integral~:
$$
C(x)= \int{\rm d}x' A(x) B(x-x') 
\eqno\neqap
$$
which in discretized form is~:
\apeqnam\disconv{disconv}
$$
C_{i} = \sum\limits_l v_l A_{l} B_{i+1-l}
\eqno\neqap
$$
where the $v_l$ are integration weights. The $v_l$ can always be arranged 
to be unity and so it is clear that \disconv) and \firstrow) are identical 
summations. This means that the multiplication of two circulant matrices
can be regarded as a convolution. The product of two circulant 
matrices can therefore be determined by multiplying the Fourier Transform (FT) of 
the first row of each of the two matrices, and then taking the inverse 
FT of this product. This yields only the first row of that product but
since it is known to be a circulant matrix the other rows are then trivially found
by shifting and wrapping around.
Similarly the inverse of a circulant matrix can be found by 
dividing the FT of the first row of the identity matrix by the FT of 
first row of the circulant matrix to be inverted, and taking the 
inverse FT.

If one has a BCCB matrix with dimensions $M_A^2 \times M_A^2$ in which 
the blocks have dimensions $M_A \times M_A$, it is fully determined by 
its first row of blocks. In equations \matmul)-\disconv) it is nowhere 
used that the individual matrix elements must be scalars. Thus two block 
circulant matrices can be multiplied by a one-dimensional convolution 
of the first row of blocks of each matrix, i.e. doing the same operation 
of convolution on every element of each block. One can visualize this 
as in diagram \stackblock{} by stacking the blocks and doing a one 
dimensional FT along columns. Since the blocks are circulant it is not 
necessary to do this operation for all columns. One does $M_A$ one-dimensional
FTs along the first rows of the blocks for each matrix and then multiplies
row by row these first rows. 

The first step is therefore to perform the $M_A$ one dimensional FTs of 
the first row of each block, for both matrices. The second step is to treat each 
of these rows as an element in an array, and the two arrays corresponding 
to matrix $A$ and $B$ are convolved. 
The two-level hierarchy of one dimensional FTs can be achieved simply by 
one two-dimensional FT, in which one direction is a horizontal one
and the other the vertical in diagram \stackblock{}. 
The two-dimensional FT is thus applied to a matrix of size $M_A \times M_A$
instead of a one dimensional FT on the first row of the full matrix. 
For a multiplication of two matrices the two-dimensional tableaux
are multiplied element by element in the Fourier domain and the 
result is inverse FT'd. The result is the first row of the BCCB product
matrix, where the other rows are obtained by shifting and wrapping around.
The inverse of a matrix in the Fourier domain is a simple division
and therefore carried out analogously to the multiplication of matrices.

By using FFT algorithms the inversion of block circulant matrices 
with circulant blocks with size $M\times M$ is thus carried out in 
${\cal O}(M^2 \log M)$ operations.

One more step is necessary in order to be able to apply this method of
inverting matrices to the problem at hand. Because of the constraint
\unitav) the block circulant matrix with circulant blocks is augmented
with one row and column. All elements of this row and column are
equal to unity except for the corner element which is $0$. Thus the 
matrix to be inverted is $A'$~:
$$
A' \equiv \left( \twomat{ A& {\bf 1}\cr {\bf 1}^{\rm T}& 0\cr} \right)
\eqno\neqap
$$
where $A$ is a BCCB matrix and ${\bf 1}$ is a column vector of which all 
$M_A^2$ elements are equal to $1$. The inverse of this partitioned matrix 
can be written as (cf.  Press \etal, 1992)~:
$$
A'^{-1} \equiv \left( \twomat{ P& Q\cr Q^{\rm T}& $-{1\over s}$\cr} \right)
\eqno\neqap
$$
in which $P$ has the same dimensions as $A$, $Q$ is a column vector, and
$s$ is a scalar~:
\apeqnam\blockpar{blockpar}
$$
\eqalign{
P &= A^{-1} - {1\over s} A^{-1}\cdot {\bf 1} \cdot {\bf 1}^{\rm T}\cdot A^{-1} \cr
Q &= {1\over s} A^{-1}\cdot {\bf 1} \cr
s &= \left( {\bf 1}^{\rm T}\cdot A^{-1}\cdot {\bf 1}\right)\cr}
\eqno\neqap
$$
Since $A$ is a block circulant matrix with circulant blocks the column vector
resulting from the product $A^{-1}\cdot {\bf 1}$ is itself a column vector 
$\alpha {\bf 1}$ where $\alpha$ is some number that depends on $A$. Using this 
\blockpar) can be simplified further~:
$$
\eqalign{
P &= \left( I - {1\over M_A^2} {\bf 1} \cdot {\bf 1}^{\rm T}\right)\cdot A^{-1} \cr
Q &= {1\over M_A^2} {\bf 1} \cr
s &= M_A^2\alpha\cr}
\eqno\neqap
$$
The matrix formed by ${\bf 1} \cdot {\bf 1}^{\rm T}$ is clearly a circulant
matrix and therefore also block circulant with circulant blocks. 
The matrix multiplications to evaluate $P$ can therefore be done 
with FTs as described above. 

The first step in this process is sectioning the first rows and rearranging.
The $M_A \times M_A$ matrix formed from sectioning and rearranging the first row 
of $I$ has only one non-zero element which is the first element of the first row
(equal to $1$). The FT of this matrix is a matrix of 
which every element is equal to $1$. The matrix formed from sectioning 
and rearranging the first row of ${1\over M_A^2} {\bf 1} \cdot {\bf 1}^{\rm T}$ 
is an $M_A \times M_A$ matrix of which every element is equal to 
${1\over M_A^2}$ and so the first element of the first row of its FT is 
equal to $1$ and all other elements are $0$.
Subtracting one from the other in the Fourier domain produces
a matrix with as the first element of the first row a $0$ and all other elements
equal to $1$. Evaluating the FT of $P$ and doing the inverse FT is then 
trivial.

All the vectors $b$ of equation \vector) collected for all pixels also
form an $M_A^2 \times M_A^2$ BCCB matrix $B$ with one extra bottom row
of $1$'s arising from the constraint \unitav). The product of this
with the $A'^{-1}$ is therefore also most easily carried out in the 
Fourier domain using $P$. Adding the final element due to the addition
of the element $Q \cdot {\bf 1}^T$ is a simple addition of $1/M_A^2$ to 
every element of $P\cdot B$. Since FTs are linear this addition can also 
be done in the Fourier domain, by setting the first element of the first 
row of $P\cdot B$ equal to $1$ before performing the inverse FT.

\bye